\documentclass[journal=nalefd,manuscript=letter]{achemso}

\usepackage[version=3]{mhchem}
\usepackage{amsmath}
\usepackage{siunitx}
\usepackage{graphicx}
\usepackage{xcolor}

\author{Valentin Fonck}
\affiliation{Institute of Condensed Matter and Nanosciences, Université catholique de Louvain (UCLouvain), 1348 Louvain-la-Neuve, Belgium}

\author{Roop K. Mech}
\affiliation{Institute of Condensed Matter and Nanosciences, Université catholique de Louvain (UCLouvain), 1348 Louvain-la-Neuve, Belgium}

\author{Mohammadali Razeghi}
\affiliation{Institute of Condensed Matter and Nanosciences, Université catholique de Louvain (UCLouvain), 1348 Louvain-la-Neuve, Belgium}

\author{Stuart Finch}
\affiliation{Physics Department, Lancaster University, Lancaster LA1 4YB, United Kingdom}

\author{Aljoscha Söll}
\affiliation{Dept. of Inorganic Chemistry, University of Chemistry and Technology Prague, Techicka 5, 166 28 Prague 6, Czech Republic}

\author{Phillip Dobson}
\affiliation{James Watt School of Engineering, University of Glasgow, Glasgow, G12 8LT, United Kingdom}

\author{Jonathan R. Weaver}
\affiliation{James Watt School of Engineering, University of Glasgow, Glasgow, G12 8LT, United Kingdom}

\author{Zdenek Sofer}
\affiliation{Dept. of Inorganic Chemistry, University of Chemistry and Technology Prague, Techicka 5, 166 28 Prague 6, Czech Republic}

\author{Oleg Kolosov}
\email{o.kolosov@lancaster.ac.uk}
\affiliation{Physics Department, Lancaster University, Lancaster LA1 4YB, United Kingdom}

\author{Jean Spièce}
\email{jean.spiece@uclouvain.be}
\affiliation{Institute of Condensed Matter and Nanosciences, Université catholique de Louvain (UCLouvain), 1348 Louvain-la-Neuve, Belgium}

\author{Pascal Gehring}
\email{pascal.gehring@uclouvain.be}
\affiliation{Institute of Condensed Matter and Nanosciences, Université Catholique de Louvain (UCLouvain), 1348 Louvain-la-Neuve, Belgium}
\affiliation{WEL Research Institute, Avenue Pasteur 6, 1300 Wavre, Belgium}

\title{Direct Nanoscale Pyroelectric Characterization of a \ce{CuInP2S6} van der Waals Nanogenerator}

\keywords{pyroelectricity, \ce{CuInP2S6}, scanning thermal microscopy, van der Waals heterostructures, 2D ferroelectrics, energy harvesting}

\begin{document}

\begin{abstract}

Pyroelectric energy conversion offers a route for harvesting time-dependent thermal fluctuations that are abundant in natural and technological environments. Two-dimensional ferroelectrics are particularly attractive for this purpose because reduced dimensionality enables ultrathin, mechanically compliant device architectures. Here, we demonstrate direct nanoscale pyroelectric characterization of an out-of-plane van der Waals nanogenerator based on \ce{CuInP2S6} (CIPS) encapsulated between few-layer graphene electrodes. A scanning thermal microscopy (SThM) probe is employed as a localized nanoscale heat source while the electrically generated response is measured in situ through the device electrodes. Harmonic detection isolates the pyroelectric signal from parasitic first-harmonic electromechanical contributions, while finite-element thermal modeling combined with probe calibration enables direct determination of the local pyroelectric coefficient from the measured electrical response. Beyond quantitative characterization, the spatially resolved measurements directly identify electrically inactive regions associated with device defects, revealing local performance-limiting features that remain hidden in conventional spatially averaged pyroelectric measurements. The presented approach establishes a versatile platform for quantitative nanoscale pyroelectric characterization and the optimization of van der Waals pyroelectric devices.

\end{abstract}

Pyroelectric energy harvesting has emerged as an attractive strategy for powering autonomous low-power electronics by converting time-dependent thermal fluctuations into electrical signals. This is especially relevant for the Internet of Things, where the large number of distributed devices creates a growing demand for compact and maintenance-free energy sources. In contrast to thermoelectric conversion, pyroelectricity does not require a static spatial temperature gradient, but instead relies on temporal changes in temperature, making it well suited to environments with fluctuating waste heat or ambient thermal cycling.

In pyroelectric materials with spontaneous polarization $P_S$, a temperature variation induces a change in polarization and therefore a measurable electrical response. The pyroelectric coefficient under constant stress can be written as

\begin{equation}
\left(\frac{dP_S}{dT}\right)_\sigma
=
p_1+p_2
=
\left(\frac{\partial P_S}{\partial T}\right)_\varepsilon
+
\sum_i
\left(\frac{\partial P_S}{\partial \varepsilon_i}\right)_T
\left(\frac{d\varepsilon_i}{dT}\right)_\sigma ,
\end{equation}

where $p_1$ is the primary pyroelectric contribution at constant strain and $p_2$ is the secondary contribution arising from thermally induced strain.\cite{born_quantum_1945,szigeti_temperature_1975,liu_mechanisms_2018} Here, $\epsilon_i$ is the strain along direction $i$ and $\sigma$ is the applied stress. A periodic temperature modulation therefore drives a periodic redistribution of surface charge that can be harvested electrically.

Two-dimensional materials provide an appealing platform for pyroelectric energy conversion. Reduced dimensionality can modify lattice dynamics, strain coupling and dielectric screening, and large pyroelectric coefficients have been predicted or reported in several layered systems, including $\beta$-\ce{In2Se3}, black phosphorene, tellurium nanosheets, \ce{CuInP2S6}, and \ce{SnSe}.\cite{jiang_giant_2022,you_room-temperature_2018,mishra_ultrahigh_2025,morozovska_strain-induced_2023,kumar_giant_2025} In addition to their intrinsic properties, van der Waals materials offer practical advantages such as mechanical flexibility, heterostructure integration, and compatibility with transparent or ultrathin electrodes. In particular, graphite or graphene electrodes provide intimate electrical contact while avoiding metallic diffusion issues that may degrade ferroelectric performance.\cite{jachalke_how_2017}

Among layered ferroelectrics, \ce{CuInP2S6} (CIPS) is particularly attractive because it exhibits out-of-plane ferroelectricity near room temperature and can be readily integrated into vertical van der Waals capacitors.\cite{belianinov_cuinp2s6_2015,zhou_van_2020} Despite considerable interest in this material system, quantitative pyroelectric characterization has primarily relied on indirect approaches based on polarization measurements. Direct measurements of the pyroelectric response are, however, essential for understanding the intrinsic performance of realistic device architectures. Moreover, pyroelectric nanogenerators are inherently heterogeneous: defects, imperfect poling, electrode interfaces and inactive regions can strongly limit the overall device performance while remaining hidden in conventional spatially averaged measurements. Access to local pyroelectric activity therefore provides both a quantitative characterization tool and a powerful diagnostic for optimizing nanoscale energy harvesters.

Existing pyroelectric characterization techniques primarily rely on laser-induced heating, either through an opaque electrode\cite{stewart_use_2009,mellinger_three-dimensional_2005,marty-dessus_space_2002,bauer_polarization_1991,lang_laserintensitymodulation_1986} or directly at the top surface of the material.\cite{caro_chromophorezeolite_1994,marlow_pyroelectric_2004,quintel_analysis_1998} While these approaches have enabled important advances, they become increasingly challenging for ultrathin or transparent electrode architectures and are ultimately limited in spatial resolution. Pyroelectric Scanning Probe Microscopy (PyroSPM) represents an important step toward nanoscale characterization by raster-scanning a conductive probe while heating is applied through the bottom electrode via absorbed laser power. Although this approach achieves nanometer-scale resolution in the measured pyrovoltage, it still requires an opaque bottom electrode and an optically transparent substrate, significantly restricting device design. Furthermore, it does not probe the local thermal coupling between the electrodes and the pyroelectric layer, a parameter that can critically influence device performance.

Here, we employ a scanning thermal microscopy (SThM) probe as a localized nanoscale heat source to directly excite pyroelectricity in a vertical CIPS-based van der Waals nanogenerator while simultaneously measuring the electrical response through the device electrodes. Harmonic analysis separates the pyroelectric signal from parasitic first-harmonic electromechanical contributions, while finite-element thermal modeling combined with probe calibration enables a direct determination of the pyroelectric coefficient from the measured electrical response. Beyond quantitative characterization, the spatially resolved measurements reveal electrically inactive regions and nanoscale defects that limit the local pyroelectric performance. The presented approach therefore establishes a versatile platform for quantitative nanoscale pyroelectric characterization and for the optimization of van der Waals pyroelectric devices.

\section{Results and Discussion}

\begin{figure*}
    \centering
    \includegraphics[width=0.98\linewidth]{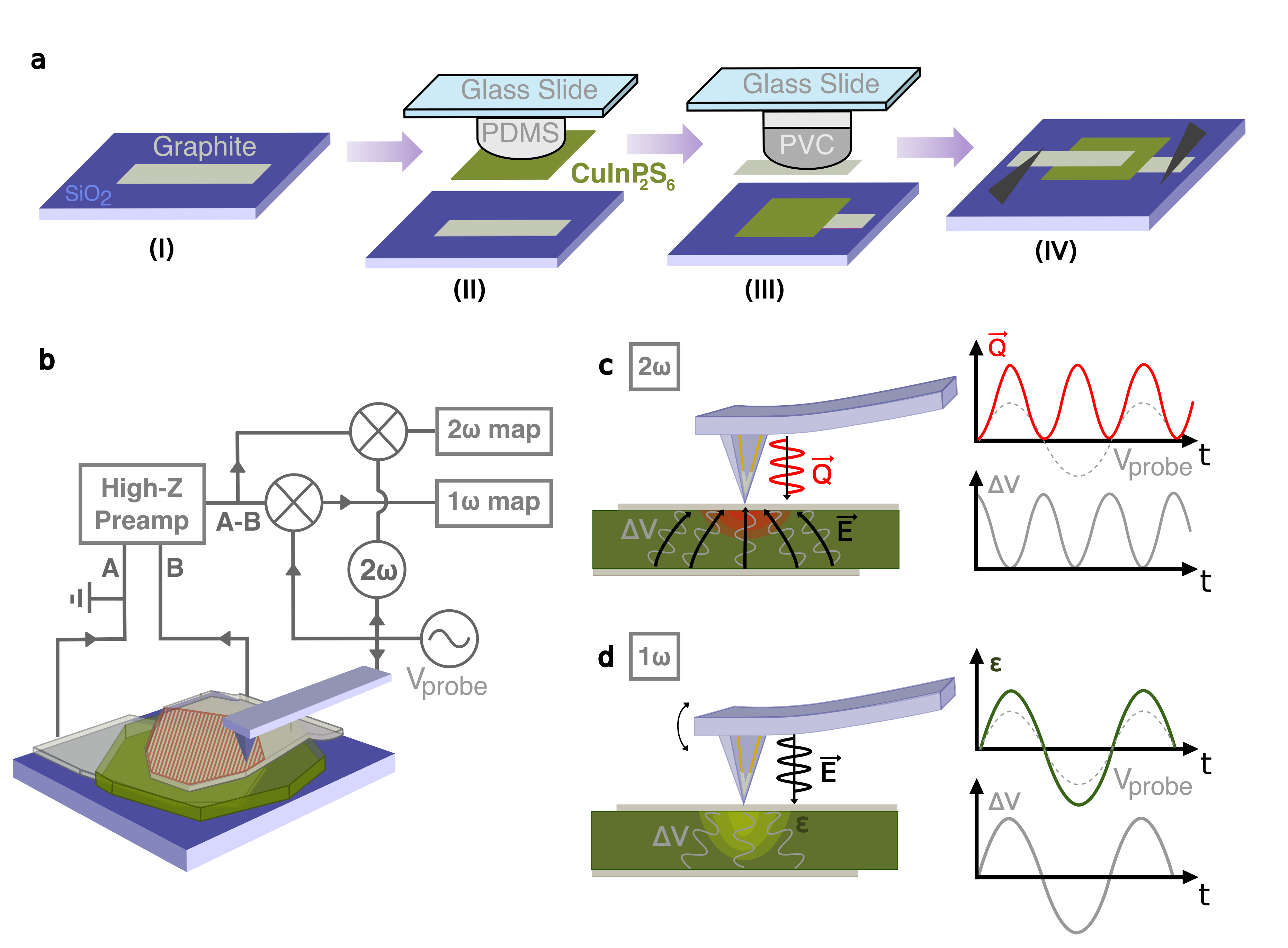}
    \caption{\textbf{a} Schematic illustration of device fabrication. (I) A bottom few-layer graphene electrode is transferred onto a \SI{285}{nm} \ce{SiO2}/Si substrate. (II) A CIPS flake is transferred onto the bottom electrode. (III) A top few-layer graphene electrode is transferred to form a vertical capacitor with an overlap region. (IV) The electrodes are contacted using indium needles and wire-bonded to a sample holder. \textbf{b} Schematic of the pyroelectric imaging setup using a scanning thermal microscopy probe as a localized heat source. Electrical signals are shown as grey solid lines while boxes represent measurement instrumentation. \textbf{c} Origin of the second harmonic signal, schematically with time-dependent vector fields within a cross-section of the tip-sample system. Schematic graphs of the heat flux $\vec{Q}$ and the measured voltage difference on the electrodes $\Delta V$ are shown as insets. \textbf{d} Similar representation for the origin of the first harmonic signal with the strain in the capacitor $\epsilon$ and the measured signal $\delta V$ as insets.}
    \label{fig:exp_scheme}
\end{figure*}

The pyroelectric device consists of a vertical van der Waals capacitor formed by a \ce{CuInP2S6} (CIPS) flake with thickness ranging from 220 to 260~nm, sandwiched between two few-layer graphene electrodes on a 285~nm thick \ce{SiO2}/Si substrate. The bottom and top electrodes (8 and 5~nm thick, respectively) define an active overlap area of $230\pm5~\mu\mathrm{m}^2$. The device is fabricated by PVC polymer transfer process \cite{mech2025versatile}, and the graphene electrodes are electrically contacted using in-situ melted indium needles,\cite{razeghi_plasmon-enhanced_2022,razeghi_single-material_2023} as illustrated in Figure~\ref{fig:exp_scheme}a (Optical micrograph in Figure~S1).

During measurement, the device is locally excited by a periodic temperature modulation generated with a scanning thermal microscopy (SThM) probe in contact with its surface (Figure~\ref{fig:exp_scheme}b). \cite{harzheim_geometrically_2018} The oscillatory heating is produced by driving the probe's integrated thermistor element with an AC voltage through a home-made Wheatstone bridge, resulting in a localized heat flux into the heterostructure. The electrical response is simultaneously measured between the top and bottom graphene electrodes using a high-impedance voltage preamplifier and lock-in detection.

The pyroelectric response is isolated using harmonic detection. An AC excitation of the SThM probe generates Joule heating proportional to $V^2$, resulting in a local temperature modulation at twice the excitation frequency ($2\omega$, Figure~\ref{fig:exp_scheme}c). Since the pyroelectric response is proportional to the rate of temperature change ($dT/dt$), the generated voltage is also detected at the second harmonic. Electrostatic and electromechanical interactions, by contrast, contribute predominantly to the first harmonic as they follow the electrostatic bending of the probe, directly proportional to the applied voltage (see Figure~\ref{fig:exp_scheme}d). This provides a direct way to isolate the pyroelectric signal from the parasitic electromechanical contributions.

\begin{figure*}
    \centering
    \includegraphics[width=0.98\linewidth]{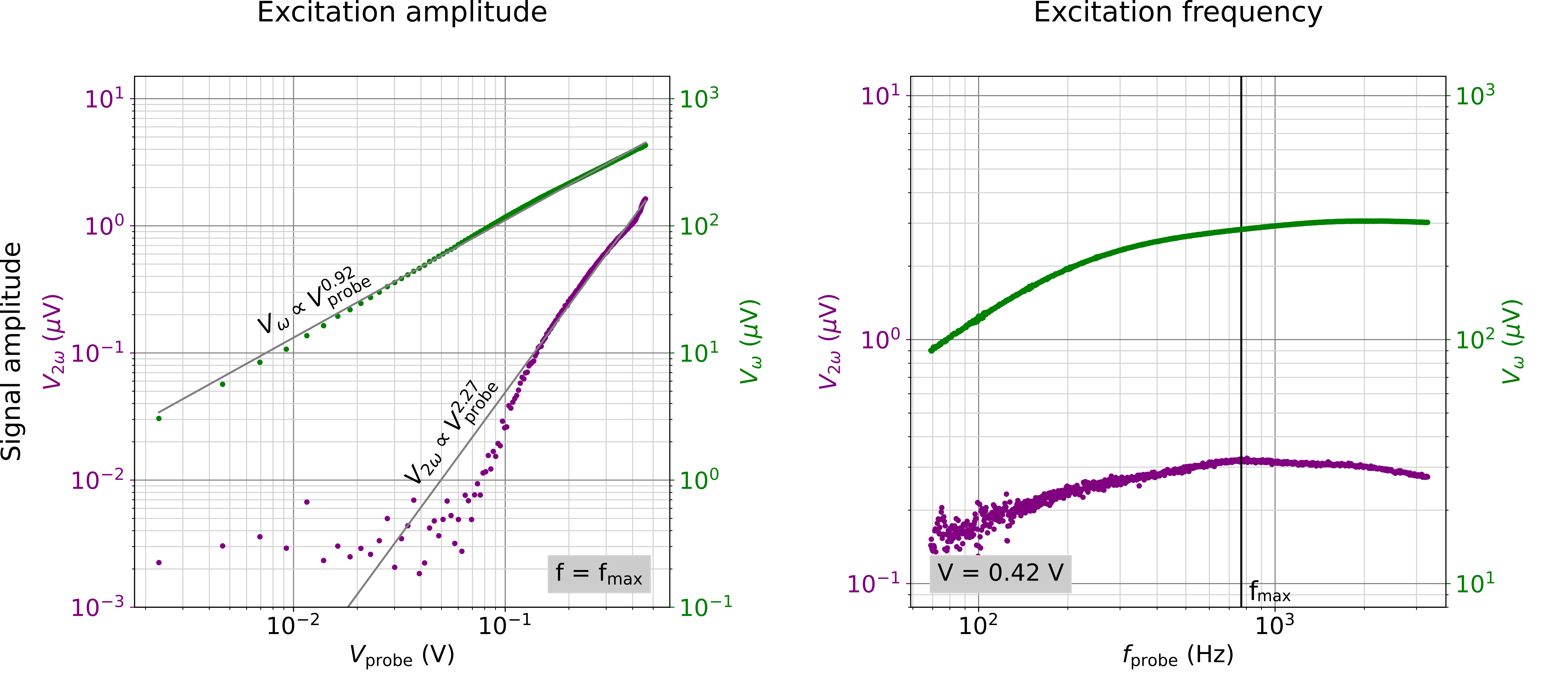}
    \caption{Parameter study on the dependence of the pyroelectric signal, shown in purple, and the electromechanical response, shown alongside in green.  \textbf{a} Signal amplitudes with respect to the excitation voltage $V_\text{probe}$. \textbf{b} Signal amplitudes with respect to the excitation frequency $f_\text{probe}$. \textbf{c} Signal phases with respect to the excitation voltage $V_\text{probe}$. \textbf{d} Signal phases with respect to the excitation frequency $f_\text{probe}$.}
    \label{fig:parametric}
\end{figure*}

This interpretation is supported by parametric measurements of the probe excitation amplitude and frequency performed on the top-electrode region (Figure~\ref{fig:parametric}). The second-harmonic response remains close to the noise floor at low excitation and then increases approximately quadratically with probe voltage (Figure~\ref{fig:parametric}a). A power-law fit yields an exponent of 2.27, consistent with the expected quadratic dependence of Joule heating. In contrast, the first-harmonic response scales approximately linearly with excitation voltage, with a power-law exponent of 0.92, as expected for electrostatic or electromechanical interactions. The corresponding phase information is shown in Figure~S2.

The frequency dependence of the second-harmonic response follows the characteristic behavior expected for localized periodic heating (Figure~\ref{fig:parametric}b). The signal vanishes in the low-frequency limit, reaches a maximum at an intermediate frequency $f_\mathrm{max}$, and decreases again once the temperature modulation becomes limited by the characteristic thermalization time of the probe--sample system. By contrast, the first-harmonic response exhibits a monotonic increase with excitation frequency, consistent with its electrostatic or electromechanical origin.
Possible contact-related artefacts were systematically investigated and excluded through dedicated control experiments (Supplementary Information~C). In particular, electrostatic interactions, friction, and water-meniscus effects cannot account for the observed second-harmonic response. Combined with the harmonic scaling, frequency dependence, phase response, and spatial localization, these controls consistently identify the measured signal as originating from local pyroelectric excitation.

\begin{figure*}
    \centering
    \includegraphics[width=0.98\linewidth]{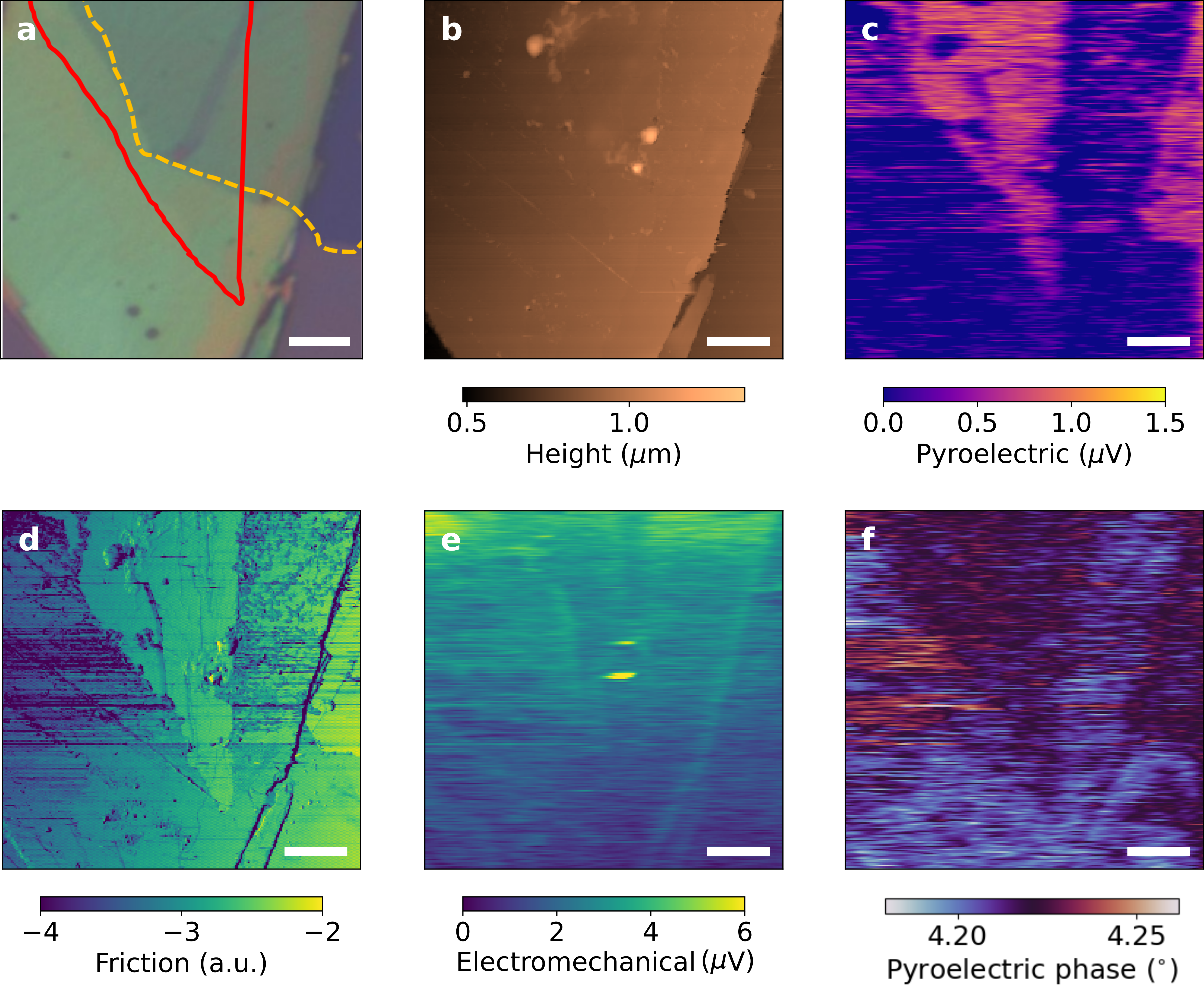}
    \caption{\textbf{a} Optical micrograph of the scanned region. The top and bottom few-layer graphene electrodes are respectively delimited with a red and yellow line. \textbf{b}Topography of the van der Waals capacitor.  \textbf{c} and \textbf{f} Amplitude and phase of the second harmonic voltage, induced by the pyroelectric effect. \textbf{d} Friction map showing mechanical contrast between the different regions. \textbf{e} Amplitude of the first harmonic voltage, induced by the electrostatic and electromechanical interactions. Scale bar: \SI{5}{\micro\meter}.}
    \label{fig:main_map}
\end{figure*}

Having established the pyroelectric origin of the second-harmonic response, we now exploit its spatial resolution to image the local pyroelectric activity of the device. The optical micrograph, topography, and friction map are shown in Figure~\ref{fig:main_map}a,b,d, respectively, with the latter providing clear mechanical contrast between the different device regions. The second-harmonic pyroelectric voltage map (Figure~\ref{fig:main_map}c) reveals a localized response that closely follows the active capacitor region defined by the overlap of the top and bottom graphene electrodes. In contrast, the first-harmonic electromechanical signal (Figure~\ref{fig:main_map}e) exhibits no significant contrast within the overlap area, highlighting the effectiveness of harmonic separation for isolating the pyroelectric response. The corresponding pyroelectric phase map (Figure~\ref{fig:main_map}f) displays a stable phase throughout the active capacitor region, further supporting the uniformity of the measured signal. While the strongest pyroelectric response is confined to the electrode overlap, a weaker signal extends over parts of the top electrode outside the active area, which is naturally explained by lateral heat spreading within the high in-plane thermal conductivity graphene electrode, as discussed below. All maps were acquired using a probe excitation of $f=\SI{769}{Hz}$ and \SI{0.46}{V_{pp}}.

We next determine the local pyroelectric coefficient from the measured electrical response. The required temperature field in the CIPS layer is obtained by combining a calibrated SThM probe response\cite{tovee_nanoscale_2012} with finite-element thermal modeling, while internal losses associated with the finite device resistance are accounted for using the high-frequency approximation justified by the measured R-C characteristics (Supplementary Information~D). Under the experimental conditions used here (\SI{0.46}{V_{pp}}), the probe calibration corresponds to an AC temperature modulation of approximately \SI{41}{K}, superimposed on a comparable DC temperature offset. The simulations reveal pronounced lateral heat spreading within the few-layer graphene electrodes, naturally explaining the finite pyroelectric response observed outside the direct electrode overlap while providing the spatial temperature distribution required for the quantitative analysis.

In an open-circuit pyroelectric capacitor, a local temperature modulation induces a local change in polarization, while the electrodes enforce a single potential difference across the entire device area. Consequently, the measured pyroelectric voltage is determined by the volume-integrated temperature field in the active CIPS layer. Under the parallel-plate approximation,

\begin{equation}
V_{\mathrm{pyro}}
=
-\frac{p}{\epsilon_0 \epsilon_r A_{\mathrm{tot}}}
\iiint_{\mathrm{CIPS}} \Delta T(x,y,z)\, dV ,
\end{equation}
where $p$ is the pyroelectric coefficient, $\epsilon_0$ and $\epsilon_r$ are the vacuum and relative permittivities, respectively, $A_{\mathrm{tot}}$ is the electrode overlap area, and $\Delta T(x,y,z)$ is the local temperature modulation in the CIPS layer obtained from finite-element analysis. Equivalently,
\begin{equation}
p
=
\epsilon_0 \epsilon_r A_{\mathrm{tot}}
\frac{|V_{\mathrm{pyro}}|}{\iiint_{\mathrm{CIPS}} \Delta T(x,y,z)\, dV}.
\end{equation}

Using the simulated temperature field (Supplementary Information~E), the measured electrode overlap area (Supplementary Information~A), and the temperature-dependent dielectric permittivity reported for CIPS,\cite{belianinov_cuinp2s6_2015} we obtain an open-circuit pyroelectric coefficient between \SI{1.91}{} and \SI{11.25}{\micro\coulomb\per\meter\squared\per\kelvin} (orders of magnitude larger than the sensitivity of our method, which, in noise-equivalent pyroelectric coefficient, can be estimated around \SI{1}{\nano\coulomb\per\meter^2\kelvin\hertz^{1/2}}). This value is approximately one order of magnitude below previously reported values for CIPS.\cite{niu_controlled_2019,morozovska_strain-induced_2023} Such a difference is, however, not unexpected, as previous studies primarily relied on theoretical or indirect approaches based on polarization measurements, whereas the present work directly probes the electrical response of an operating van der Waals capacitor under localized thermal excitation. The measured coefficient therefore inherently includes the influence of interfaces, finite thickness, electrode coupling, and nonuniform thermal transport.

A second contributing factor is the finite thickness of the CIPS layer. With a thickness of approximately \SI{220}{\nano\meter}, the device approaches the critical regime where CIPS evolves from a single ferroelectric domain toward a paraelectric configuration,\cite{belianinov_cuinp2s6_2015} which is expected to reduce the pyroelectric response. The reported range is dominated by the strong temperature dependence of the dielectric permittivity and the resulting uncertainty in the local base temperature beneath the heated probe.\cite{tovee_nanoscale_2012} Future implementations combining independent DC and AC probe-heating control,\cite{menges_nanoscale_2016,harnack_scanning_2025,fonck_characterization_2026} together with systematic thickness-dependent studies, should enable even more accurate determination of the intrinsic pyroelectric coefficient and identify the optimum trade-off between ferroelectric stability and thermal device performance. 

\begin{figure}[h!]
    \centering
    \includegraphics[width=0.8\linewidth]{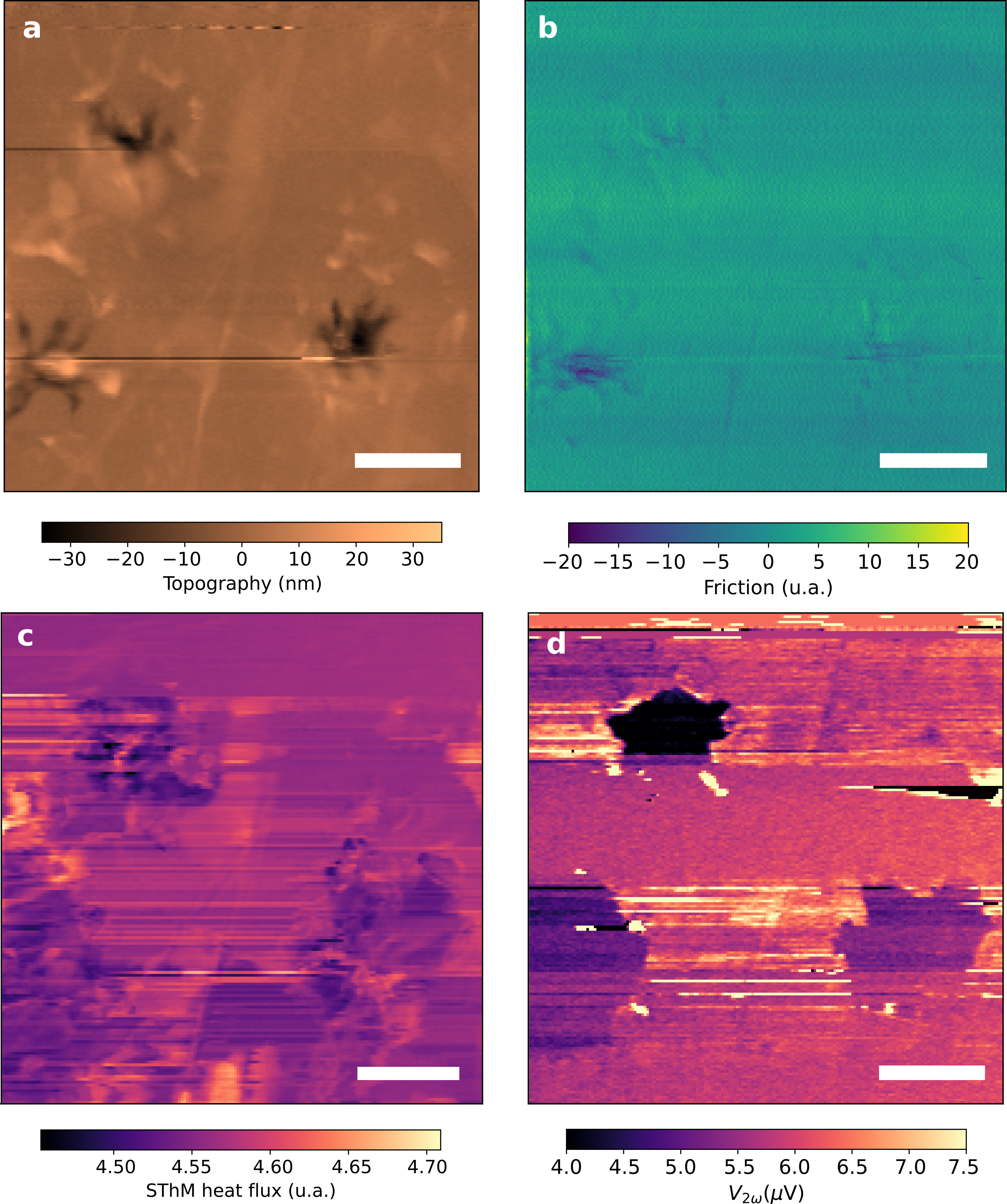}
    \caption{\textbf{a} Topography of defects in the pyrogenerator's top electrode. \textbf{b} Lateral friction map. \textbf{c} SThM heat-flux signal, measured in arbitrary units. \textbf{d} Pyrovoltage amplitude measured between the two electrodes. Scale bar: \SI{400}{\nano\meter}.}
    \label{fig:holes}
\end{figure}

Beyond quantitative determination of the pyroelectric coefficient, the spatial resolution of the technique enables direct identification of locally inactive regions within operating pyroelectric devices. Defects in the ferroelectric material or the electrodes can strongly degrade the performance of pyroelectric generators and detectors while remaining hidden in spatially averaged measurements. As a proof of concept, we image defects in the top graphene electrode, potentially originating from repeated scans or internal stress during fabrication. Figure~\ref{fig:holes} compares four simultaneously acquired imaging channels. The topography, shown in Figure~\ref{fig:holes}a, reveals holes surrounded by wrinkled graphene, whereas the lateral friction map, in Figure~\ref{fig:holes}b, shows only weak contrast and the SThM heat-flux channel exhibits a response affected by contact artefacts in Figure~\ref{fig:holes}c. In contrast, the pyroelectric voltage map, shown in Figure~\ref{fig:holes}d, displays sharply defined inactive regions, directly identifying electrically disconnected areas of the device. These results highlight the unique capability of local pyroelectric imaging to reveal performance-limiting defects that cannot be unambiguously identified by topography, friction, or thermal conductance measurements alone.

\section{Conclusion}

In summary, we demonstrate direct nanoscale pyroelectric characterization of a vertical \ce{CuInP2S6} van der Waals nanogenerator using a scanning thermal microscopy probe as a localized heat source. Harmonic detection combined with calibrated thermal modeling enables direct determination of the local pyroelectric coefficient from the electrical response of an operating device, while the spatial resolution simultaneously reveals electrically inactive regions that remain hidden in conventional pyroelectric measurements. By combining quantitative pyroelectric characterization with local defect identification, the presented approach provides a versatile platform for the development and optimization of pyroelectric van der Waals heterostructures and other nanoscale thermal energy conversion devices.

\section{Methods}

\textbf{Scanning probe microscopy}
A KNT-STHM probe was used in this work, it features a a Pd thermistor at the apex of a silicon nitride cantilever. The nominal tip’s radius is 100 nm. \cite{dobson_new_2007} The scanning was performed in constant force mode and in air on a Bruker Dimension Icon AFM set-up. Unless specified otherwise, the scanning speed was set to \SI{6}{\micro\meter\per\second}. An AC voltage generated by a lock-in amplifier (Zurich Instruments HF2LI) was applied to the probe thermistor through a homemade AC Wheatstone bridge, producing periodic Joule heating of the tip. The resulting electrical response of the device was measured in voltage mode using a high-impedance preamplifier (Brookdeal 5004). The preamplifier output was demodulated using the lock-in amplifier, allowing analysis of the first- and second-harmonic responses.

\textbf{Material synthesis.} \ce{CuInP2S6} (CIPS) is a layered van der Waals compound composed of Cu and In cations embedded in a \ce{(P2S6)^{4-}} framework. Below its Curie temperature, the displacement of Cu ions breaks inversion symmetry and gives rise to out-of-plane ferroelectricity.\cite{zhou_van_2020} In its stoichiometric form, CIPS is a wide-bandgap semiconductor.

Single crystals of \ce{CuInP2S6} were grown by chemical vapor transport (CVT) in sealed quartz ampoules. Quartz ampoules (40 $\times$ 250~mm) were loaded with a stoichiometric mixture of Cu (99.9\%, $-100$ mesh, Alfa Aesar, Germany), In (99.99\%, $-100$ mesh, Alfa Aesar, Germany), S (99.9999\%, 2--6~mm, Wuhan Xinrong New Materials Co., China), and P (99.9999\%, 2--6~mm, Wuhan Xinrong New Materials Co., China), corresponding to 20~g of \ce{CuInP2S6}. The ampoules were sealed under high vacuum ($<1\times10^{-3}$~Pa) using an oil diffusion pump and an oxygen--hydrogen torch. Iodine (0.7~g, 99.9\%, granules, Fisher Scientific, UK) was added as the transport agent together with a 2~at.\% excess of sulfur and phosphorus.

The sealed ampoules were first heated in a muffle furnace to react the elemental precursors using a heating/cooling rate of \SI{1}{\celsius\per\minute}: \SI{450}{\celsius} for 25~h, \SI{500}{\celsius} for 50~h, and finally \SI{600}{\celsius} for 50~h. The ampoules were subsequently transferred to a two-zone furnace for crystal growth. Initially, the growth zone was heated to \SI{800}{\celsius} while the source zone was maintained at \SI{600}{\celsius}. After two days, the thermal gradient was reversed and maintained for 14~days, with the growth zone at \SI{650}{\celsius} and the source zone at \SI{750}{\celsius}. After completion of the CVT growth, the ampoules were cooled to room temperature and opened inside an argon-filled glovebox.

\textbf{Device fabrication.}
CIPS flakes were mechanically exfoliated onto polydimethylsiloxane (PDMS). Few-layer graphene flakes with thicknesses ranging from 3 to 10~nm were exfoliated onto a \ce{SiO2}/Si substrate. A selected CIPS flake (260~nm thick) was transferred onto a bottom few-layer graphene flake to form the active dielectric layer. A second few-layer graphene flake (5~nm thick) was subsequently transferred onto the CIPS using a PVC polymer pickup-and-release method under ambient conditions.\cite{mech2025versatile} Care was taken to avoid electrical shorting between the two electrodes. In the final fabrication step, the few-layer graphene electrodes were contacted using indium needles melted at approximately \SI{140}{\celsius}, and the contacts were wire-bonded to a standard sample holder.

\textbf{Thermal modeling.}
Finite-element simulations were used to determine the local temperature modulation within the heterostructure and the effective heating area beneath the SThM tip. The probe temperature was estimated from literature calibration curves.\cite{tovee_nanoscale_2012} A detailed description of the model, underlying assumptions, and simulation methodology is provided in Supplementary Information~E.

\begin{acknowledgement}

The authors acknowledge financial support from the F.R.S.-FNRS of Belgium (FNRS-CQ-1.C044.21-SMARD, FNRS-CDR-J.0068.21-SMARD, FNRS-MIS-F.4523.22-TopoBrain, FNRS-PDR-T.0128.24-ART-MULTI, FNRS-CR-1.B.463.22-MouleFrits, FNRS-FRIA-1.E092.23-TOTEM), from the EU (ERC-StG-10104144-MOUNTAIN), from the Fédération Wallonie--Bruxelles through ARC Grant No.~21/26-116, and from the FWO and F.R.S.-FNRS under the Excellence of Science (EOS) programme (40007563-CONNECT). Z.S. acknowledges support from the ERC-CZ programme (project LL2101) funded by the Ministry of Education, Youth and Sports (MEYS) of the Czech Republic, and from the Advanced Multiscale Materials for Key Enabling Technologies project (No.~CZ.02.01.01/00/22\_008/0004558), co-funded by the European Union.

\end{acknowledgement}

\begin{suppinfo}
Additional experimental data, discussions and details on the methods used are provided in the Supporting Information.
\end{suppinfo}

\bibliography{mybib}

@article{born_quantum_1945,
	title = {On the {Quantum} {Theory} of {Pyroelectricity}},
	volume = {17},
	doi = {10.1103/RevModPhys.17.245},
	number = {2-3},
	journal = {Reviews of Modern Physics},
	author = {Born, Max},
	year = {1945},
	pages = {245--251},
}

@article{szigeti_temperature_1975,
	title = {Temperature {Dependence} of {Pyroelectricity}},
	volume = {35},
	doi = {10.1103/PhysRevLett.35.1532},
	number = {22},
	journal = {Physical Review Letters},
	author = {Szigeti, B.},
	year = {1975},
	pages = {1532--1534}
}

@article{liu_mechanisms_2018,
	title = {Mechanisms of {Pyroelectricity} in {Three}- and {Two}-{Dimensional} {Materials}},
	volume = {120},
	doi = {10.1103/PhysRevLett.120.207602},
	number = {20},
	journal = {Physical Review Letters},
	author = {Liu, Jian},
	year = {2018},
}

@article{jiang_giant_2022,
	title = {Giant pyroelectricity in nanomembranes},
	volume = {607},
	copyright = {2022 The Author(s), under exclusive licence to Springer Nature Limited},
	issn = {1476-4687},
	url = {https://www.nature.com/articles/s41586-022-04850-7},
	doi = {10.1038/s41586-022-04850-7},
	language = {en},
	number = {7919},
	journal = {Nature},
	author = {Jiang, Jie and Zhang, Lifu and Ming, Chen and Zhou, Hua and Bose, Pritom and Guo, Yuwei and Hu, Yang and Wang, Baiwei and Chen, Zhizhong and Jia, Ru and Pendse, Saloni and Xiang, Yu and Xia, Yaobiao and Lu, Zonghuan and Wen, Xixing and Cai, Yao and Sun, Chengliang and Wang, Gwo-Ching and Lu, Toh-Ming and Gall, Daniel and Sun, Yi-Yang and Koratkar, Nikhil and Fohtung, Edwin and Shi, Yunfeng and Shi, Jian},
	month = jul,
	year = {2022},
	pages = {480--485},
}

@article{you_room-temperature_2018,
	title = {Room-temperature pyro-catalytic hydrogen generation of {2D} few-layer black phosphorene under cold-hot alternation},
	volume = {9},
	copyright = {2018 The Author(s)},
	issn = {2041-1723},
	doi = {10.1038/s41467-018-05343-w},
	language = {en},
	number = {1},
	urldate = {2025-10-14},
	journal = {Nature Communications},
	author = {You, Huilin and Jia, Yanmin and Wu, Zheng and Wang, Feifei and Huang, Haitao and Wang, Yu},
	month = jul,
	year = {2018},
	note = {Publisher: Nature Publishing Group},
	pages = {2889},
}

@article{niu_controlled_2019,
	title = {Controlled synthesis and room-temperature pyroelectricity of {CuInP2S6} ultrathin flakes},
	volume = {58},
	issn = {2211-2855},
	doi = {10.1016/j.nanoen.2019.01.085},
	journal = {Nano Energy},
	author = {Niu, Lin and Liu, Fucai and Zeng, Qingsheng and Zhu, Xiaoyang and Wang, Yanlong and Yu, Peng and Shi, Jia and Lin, Junhao and Zhou, Jiadong and Fu, Qundong and Zhou, Wu and Yu, Ting and Liu, Xinfeng and Liu, Zheng},
	month = apr,
	year = {2019},
	pages = {596--603},
}

@article{kumar_giant_2025,
	title = {Giant {Pyroelectric} {Figure} of {Merits} in {Strain}-{Engineered} {Ferroelectric} {2D}-{SnSe} {Layered} {Nanosheets}: {An} {Efficient} {Transient} {Thermal} {Energy} {Harvester}},
	volume = {19},
	issn = {1936-0851},
	shorttitle = {Giant {Pyroelectric} {Figure} of {Merits} in {Strain}-{Engineered} {Ferroelectric} {2D}-{SnSe} {Layered} {Nanosheets}},
	number = {20},
	journal = {ACS Nano},
	author = {Kumar, Ajay and Jain, Ayushi and Naskar, Sudip and Ram, Shanker and Bera, Chandan and Mandal, Dipankar},
	month = may,
	year = {2025},
	pages = {19373--19383},
}

@article{mishra_ultrahigh_2025,
	title = {Ultrahigh pyroelectricity in monoelemental two-dimensional tellurium},
	volume = {111},
	doi = {10.1103/PhysRevB.111.155436},
	number = {15},
	journal = {Physical Review B},
	author = {Mishra, Hari Krishna and Jain, Ayushi and Saini, Dalip and Mondal, Bidya and Bera, Chandan and Ram, Shanker and Mandal, Dipankar},
	month = apr,
	year = {2025},
	pages = {155436},
}

@article{mech2025versatile,
  title={Versatile polymer method to dry-flip two-dimensional moir{\'e} heterostructures for nanoscale surface characterization},
  author={Mech, Roop K and Spi{\`e}ce, Jean and Watanabe, Kenji and Taniguchi, Takashi and Kesavan, Bhagyanath Paliyottil and Sofer, Zden{\v{e}}k and Gehring, Pascal},
  journal={Physical Review B},
  volume={111},
  number={19},
  pages={195406},
  year={2025},
  publisher={APS}
}

@article{tovee_nanoscale_2012,
	title = {Nanoscale spatial resolution probes for scanning thermal microscopy of solid state materials},
	volume = {112},
	issn = {0021-8979},
	url = {https://doi.org/10.1063/1.4767923},
	doi = {10.1063/1.4767923},
	number = {11},
	journal = {Journal of Applied Physics},
	author = {Tovee, P. and Pumarol, M. and Zeze, D. and Kjoller, Kevin and Kolosov, O.},
	month = dec,
	year = {2012},
	pages = {114317},
}

@article{zhou_van_2020,
	title = {Van der {Waals} layered ferroelectric {CuInP2S6}: {Physical} properties and device applications},
	volume = {16},
	issn = {2095-0470},
	shorttitle = {Van der {Waals} layered ferroelectric {CuInP2S6}},
	doi = {10.1007/s11467-020-0986-0},
	language = {en},
	number = {1},
	urldate = {2026-01-22},
	journal = {Frontiers of Physics},
	author = {Zhou, Shuang and You, Lu and Zhou, Hailin and Pu, Yong and Gui, Zhigang and Wang, Junling},
	month = sep,
	year = {2020},
	pages = {13301},
}

@article{xu_dynamic_2018,
	title = {Dynamic {Interfacial} {Mechanical}-{Thermal} {Characteristics} of {Atomically} {Thin} {Two}-{Dimensional} {Crystals}},
	volume = {10},
	doi = {10.1039/C8NR03586E},
	journal = {Nanoscale},
	author = {Xu, Kunqi and Ye, Shili and Lei, Le and Meng, Lan and Hussain, Sabir and Zheng, Zhiyue and Zeng, Huarong and Ji, Wei and Xu, Rui and Cheng, Zhihai},
	month = jun,
	year = {2018},
}

@article{kim_ultra-high_2012,
	title = {Ultra-{High} {Vacuum} {Scanning} {Thermal} {Microscopy} for {Nanometer} {Resolution} {Quantitative} {Thermometry}},
	volume = {6},
	issn = {1936-0851},
	url = {https://doi.org/10.1021/nn300774n},
	doi = {10.1021/nn300774n},
	number = {5},
	urldate = {2026-02-10},
	journal = {ACS Nano},
	author = {Kim, Kyeongtae and Jeong, Wonho and Lee, Woochul and Reddy, Pramod},
	month = may,
	year = {2012},
	note = {Publisher: American Chemical Society},
	pages = {4248--4257},
}

@article{mcclelland_signal_2025,
	title = {Signal size and resolution of scanning thermal microscopy in air and vacuum},
	volume = {15},
	copyright = {2025 This is a U.S. Government work and not under copyright protection in the US; foreign copyright protection may apply},
	issn = {2045-2322},
	url = {https://www.nature.com/articles/s41598-025-95648-w},
	doi = {10.1038/s41598-025-95648-w},
	language = {en},
	number = {1},
	urldate = {2026-02-10},
	journal = {Scientific Reports},
	author = {McClelland, Jabez J. and Strelcov, Evgheni and Chand, Ami},
	month = apr,
	year = {2025},
	note = {Publisher: Nature Publishing Group},
	pages = {11142},
}

@article{spiece_improving_2018,
	title = {Improving accuracy of nanothermal measurements via spatially distributed scanning thermal microscope probes},
	volume = {124},
	issn = {0021-8979},
	url = {https://doi.org/10.1063/1.5031085},
	doi = {10.1063/1.5031085},
	number = {1},
	urldate = {2026-02-10},
	journal = {Journal of Applied Physics},
	author = {Spiece, J. and Evangeli, C. and Lulla, K. and Robson, A. and Robinson, B. and Kolosov, O.},
	month = jul,
	year = {2018},
	pages = {015101},
}

@article{belianinov_cuinp2s6_2015,
	title = {{CuInP2S6} {Room} {Temperature} {Layered} {Ferroelectric}},
	volume = {15},
	issn = {1530-6984},
	url = {https://doi.org/10.1021/acs.nanolett.5b00491},
	doi = {10.1021/acs.nanolett.5b00491},
	number = {6},
	urldate = {2026-02-11},
	journal = {Nano Letters},
	author = {Belianinov, A. and He, Q. and Dziaugys, A. and Maksymovych, P. and Eliseev, E. and Borisevich, A. and Morozovska, A. and Banys, J. and Vysochanskii, Y. and Kalinin, S. V.},
	month = jun,
	year = {2015},
	note = {Publisher: American Chemical Society},
	pages = {3808--3814},
}

@article{jachalke_how_2017,
	title = {How to measure the pyroelectric coefficient?},
	volume = {4},
	issn = {1931-9401},
	url = {https://doi.org/10.1063/1.4983118},
	doi = {10.1063/1.4983118},
	number = {2},
	urldate = {2026-02-09},
	journal = {Applied Physics Reviews},
	author = {Jachalke, S. and Mehner, E. and Stöcker, H. and Hanzig, J. and Sonntag, M. and Weigel, T. and Leisegang, T. and Meyer, D. C.},
	month = may,
	year = {2017},
	pages = {021303},
}

@article{zhang_impact_2020,
	title = {Impact of {Leakage} for {Electricity} {Generation} by {Pyroelectric} {Converter}},
	volume = {14},
	url = {https://link.aps.org/doi/10.1103/PhysRevApplied.14.064079},
	doi = {10.1103/PhysRevApplied.14.064079},
	number = {6},
	urldate = {2025-08-28},
	journal = {Physical Review Applied},
	author = {Zhang, Chenbo and Zeng, Zhuohui and Zhu, Zeyuan and Karami, Mostafa and Chen, Xian},
	month = dec,
	year = {2020},
	note = {Publisher: American Physical Society},
	pages = {064079},
}

@article{razeghi_plasmon-enhanced_2022,
	title = {Plasmon-enhanced photoresponse of single silver nanowires and their network devices},
	volume = {7},
	url = {https://pubs.rsc.org/en/content/articlelanding/2022/nh/d1nh00629k},
	doi = {10.1039/D1NH00629K},
	language = {en},
	number = {4},
	urldate = {2026-03-19},
	journal = {Nanoscale Horizons},
	author = {Razeghi, Mohammadali and Üstünçelik, Merve and Shabani, Farzan and Volkan Demir, Hilmi and Serkan Kasırga, T.},
	year = {2022},
	note = {Publisher: Royal Society of Chemistry},
	pages = {396--402},
}

@article{razeghi_single-material_2023,
	title = {Single-material {MoS2} thermoelectric junction enabled by substrate engineering},
	volume = {7},
	copyright = {2023 The Author(s)},
	issn = {2397-7132},
	url = {https://www.nature.com/articles/s41699-023-00406-z},
	doi = {10.1038/s41699-023-00406-z},
	language = {en},
	number = {1},
	urldate = {2026-03-19},
	journal = {npj 2D Materials and Applications},
	author = {Razeghi, Mohammadali and Spiece, Jean and Oğuz, Oğuzhan and Pehlivanoğlu, Doruk and Huang, Yubin and Sheraz, Ali and Başçı, Uğur and Dobson, Phillip S. and Weaver, Jonathan M. R. and Gehring, Pascal and Kasırga, T. Serkan},
	month = may,
	year = {2023},
	note = {Publisher: Nature Publishing Group},
	pages = {36},
}

@article{fonck_characterization_2026,
	title = {Characterization of {Heat} {Transfer} in 3-{D} {CMOS} {Structures} {Using} {Sideband} {Scanning} {Thermal} {Wave} {Microscopy}},
	volume = {75},
	issn = {1557-9662},
	url = {https://ieeexplore.ieee.org/abstract/document/11435171},
	doi = {10.1109/TIM.2026.3674275},
	urldate = {2026-04-23},
	journal = {IEEE Transactions on Instrumentation and Measurement},
	author = {Fonck, Valentin and Razeghi, Mohammadali and Spièce, Jean and Dobson, Phillip and Weaver, Jonathan and Ridgard, George and Noah, Grayson M. and Gehring, Pascal},
	year = {2026},
	pages = {1--9},
}

@article{menges_nanoscale_2016,
	title = {Nanoscale thermometry by scanning thermal microscopy},
	volume = {87},
	issn = {0034-6748},
	url = {https://doi.org/10.1063/1.4955449},
	doi = {10.1063/1.4955449},
	number = {7},
	urldate = {2026-03-16},
	journal = {Review of Scientific Instruments},
	author = {Menges, Fabian and Riel, Heike and Stemmer, Andreas and Gotsmann, Bernd},
	month = jul,
	year = {2016},
	pages = {074902},
}

@article{harnack_scanning_2025,
	title = {Scanning {Thermal} {Microscopy} {Method} for {Self}-{Heating} in {Nonlinear} {Devices} and {Application} to {Filamentary} {Resistive} {Random}-{Access} {Memory}},
	volume = {19},
	issn = {1936-0851},
	url = {https://doi.org/10.1021/acsnano.4c12784},
	doi = {10.1021/acsnano.4c12784},
	number = {5},
	urldate = {2025-06-18},
	journal = {ACS Nano},
	author = {Harnack, Nele and Rodehutskors, Sophie and Gotsmann, Bernd},
	month = feb,
	year = {2025},
	note = {Publisher: American Chemical Society},
	pages = {5342--5352},
}

@article{morozovska_strain-induced_2023,
	title = {The strain-induced transitions of the piezoelectric, pyroelectric, and electrocaloric properties of the {CuInP2S6} films},
	volume = {13},
	issn = {2158-3226},
	url = {https://doi.org/10.1063/5.0178854},
	doi = {10.1063/5.0178854},
	number = {12},
	urldate = {2025-09-24},
	journal = {AIP Advances},
	author = {Morozovska, Anna N. and Eliseev, Eugene A. and Yurchenko, Lesya P. and Laguta, Valentyn V. and Liu, Yongtao and Kalinin, Sergei V. and Kholkin, Andrei L. and Vysochanskii, Yulian M.},
	month = dec,
	year = {2023},
	pages = {125306},
}

@article{mellinger_three-dimensional_2005,
	title = {Three-dimensional mapping of polarization profiles with thermal pulses},
	volume = {86},
	issn = {0003-6951},
	url = {https://pubs.aip.org/aip/apl/article/86/8/082903/328848/Three-dimensional-mapping-of-polarization-profiles},
	doi = {10.1063/1.1870124},
	language = {en},
	number = {8},
	urldate = {2026-04-10},
	journal = {Applied Physics Letters},
	author = {Mellinger, Axel and Singh, Rajeev and Wegener, Michael and Wirges, Werner and Gerhard-Multhaupt, Reimund and Lang, Sidney B.},
	month = feb,
	year = {2005},
	note = {Publisher: AIP Publishing},
}

@article{bauer_polarization_1991,
	title = {Polarization distribution of thermally poled {PVDF} films, measured with a heat wave method ({LIMM})},
	volume = {118},
	issn = {0015-0193},
	url = {https://doi.org/10.1080/00150199108014772},
	doi = {10.1080/00150199108014772},
	number = {1},
	urldate = {2026-04-10},
	journal = {Ferroelectrics},
	author = {Bauer, S. and Ploss, B.},
	month = jun,
	year = {1991},
	note = {Publisher: Taylor \& Francis
\_eprint: https://doi.org/10.1080/00150199108014772},
	pages = {363--378},
}

@article{stewart_use_2009,
	title = {Use of scanning {LIMM} ({Laser} {Intensity} {Modulation} {Method}) to characterise polarisation variability in dielectric materials},
	volume = {183},
	issn = {1742-6596},
	url = {https://doi.org/10.1088/1742-6596/183/1/012001},
	doi = {10.1088/1742-6596/183/1/012001},
	language = {en},
	number = {1},
	urldate = {2026-04-10},
	journal = {Journal of Physics: Conference Series},
	author = {Stewart, Mark and Cain, Markys},
	month = aug,
	year = {2009},
	pages = {012001},
}

@article{lang_laserintensitymodulation_1986,
	title = {Laser‐intensity‐modulation method: {A} technique for determination of spatial distributions of polarization and space charge in polymer electrets},
	volume = {59},
	issn = {0021-8979},
	shorttitle = {Laser‐intensity‐modulation method},
	url = {https://doi.org/10.1063/1.336352},
	doi = {10.1063/1.336352},
	number = {6},
	urldate = {2026-04-10},
	journal = {Journal of Applied Physics},
	author = {Lang, Sidney B. and Das‐Gupta, D. K.},
	month = mar,
	year = {1986},
	pages = {2151--2160},
}

@article{marty-dessus_space_2002,
	title = {Space charge cartography by {FLIMM}: a three-dimensional approach},
	volume = {35},
	issn = {0022-3727},
	shorttitle = {Space charge cartography by {FLIMM}},
	url = {https://doi.org/10.1088/0022-3727/35/24/316},
	doi = {10.1088/0022-3727/35/24/316},
	language = {en},
	number = {24},
	urldate = {2026-04-10},
	journal = {Journal of Physics D: Applied Physics},
	author = {Marty-Dessus, D. and Berquez, L. and Petre, A. and Franceschi, J. L.},
	month = nov,
	year = {2002},
	pages = {3249},
}

@article{caro_chromophorezeolite_1994,
	title = {Chromophore–zeolite composites: {The} organizing role of molecular sieves},
	volume = {6},
	copyright = {Copyright © 1994 Verlag GmbH \& Co. KGaA, Weinheim},
	issn = {1521-4095},
	shorttitle = {Chromophore–zeolite composites},
	url = {https://onlinelibrary.wiley.com/doi/abs/10.1002/adma.19940060517},
	doi = {10.1002/adma.19940060517},
	number = {5},
	urldate = {2026-04-10},
	journal = {Advanced Materials},
	author = {Caro, JÜRgen and Marlow, Frank and Wübbenhorst, Michael},
	year = {1994},
	note = {\_eprint: https://advanced.onlinelibrary.wiley.com/doi/pdf/10.1002/adma.19940060517},
	pages = {413--416},
}

@article{marlow_pyroelectric_2004,
	title = {Pyroelectric {Effects} on {Molecular} {Sieve} {Crystals} {Loaded} with {Dipole} {Molecules}},
	volume = {6},
	doi = {10.1021/j100098a029},
	journal = {Advanced Materials},
	author = {Marlow, Frank and Wübbenhorst, Michael and Caro, J.},
	month = apr,
	year = {2004},
	pages = {413--416},
}

@article{quintel_analysis_1998,
	title = {Analysis of the {Polarization} {Distribution} in a {Polar} {Perhydrotriphenylene} {Inclusion} {Compound} by {Scanning} {Pyroelectric} {Microscopy}},
	volume = {102},
	issn = {1520-6106, 1520-5207},
	url = {https://pubs.acs.org/doi/10.1021/jp973350p},
	doi = {10.1021/jp973350p},
	language = {en},
	number = {22},
	urldate = {2026-04-10},
	journal = {The Journal of Physical Chemistry B},
	author = {Quintel, Andrea and Hulliger, Jürg and Wübbenhorst, Michael},
	month = may,
	year = {1998},
	pages = {4277--4283},
}

@inproceedings{dobson_new_2007,
	title = {New {Methods} for {Calibrated} {Scanning} {Thermal} {Microscopy} ({SThM})},
	url = {https://ieeexplore.ieee.org/abstract/document/4388498},
	doi = {10.1109/ICSENS.2007.4388498},
	urldate = {2025-10-22},
	booktitle = {2007 {IEEE} {SENSORS}},
	author = {Dobson, Phillip S. and Weaver, John M. R. and Mills, Gordon},
	month = oct,
	year = {2007},
	note = {ISSN: 1930-0395},
	pages = {708--711},
}

@article{harzheim_geometrically_2018,
	title = {Geometrically {Enhanced} {Thermoelectric} {Effects} in {Graphene} {Nanoconstrictions}},
	volume = {18},
	issn = {1530-6984},
	url = {https://doi.org/10.1021/acs.nanolett.8b03406},
	doi = {10.1021/acs.nanolett.8b03406},
	number = {12},
	urldate = {2025-06-18},
	journal = {Nano Letters},
	publisher = {American Chemical Society},
	author = {Harzheim, Achim and Spiece, Jean and Evangeli, Charalambos and McCann, Edward and Falko, Vladimir and Sheng, Yuewen and Warner, Jamie H. and Briggs, G. Andrew D. and Mol, Jan A. and Gehring, Pascal and Kolosov, Oleg V.},
	month = dec,
	year = {2018},
	pages = {7719--7725},
}

\end{document}


\section{Supplementary Material for "Direct measurement of the pyroelectric effect in a Van der Waals nanogenerator"}

\subsection{Estimation of electrode's surface}
\begin{figure}[h!]
    \centering
    \includegraphics[width=0.85\linewidth]{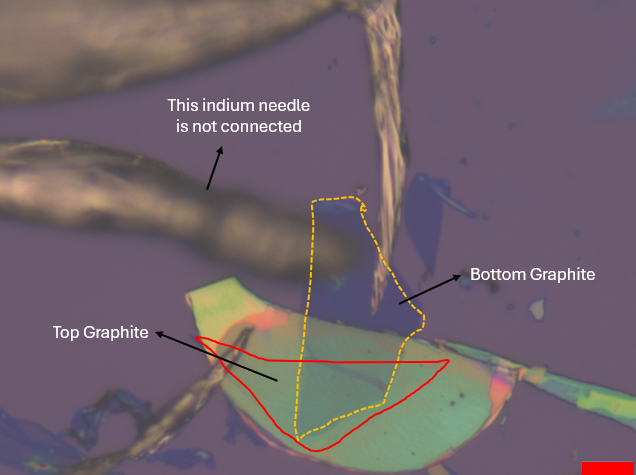}
    \caption{Micrograph of the sample presented in the main text. The regions of the top and bottom electrodes are highlighted with colored lines. Scale bar : 10~$\mu$m}
    \label{fig:si_electrode_surface}
\end{figure}

Overlaying microscope images of the stack at the different fabrication steps, illustrated in the main text at Figure~1a, allows to estimate the are of overlap between the electrodes. Figure~\ref{fig:si_electrode_surface} shows the final device with the two few layers graphene electrodes, the CIPS flake and the indium needles. Integrating over the surface of the overlapped region yields a surface of approximately 230~$\mu$m${}^2$.

\clearpage

\subsection{Phase of the pyroelectric and electromechanical signals}
\begin{figure*}
    \centering
    \includegraphics[width=0.98\linewidth]{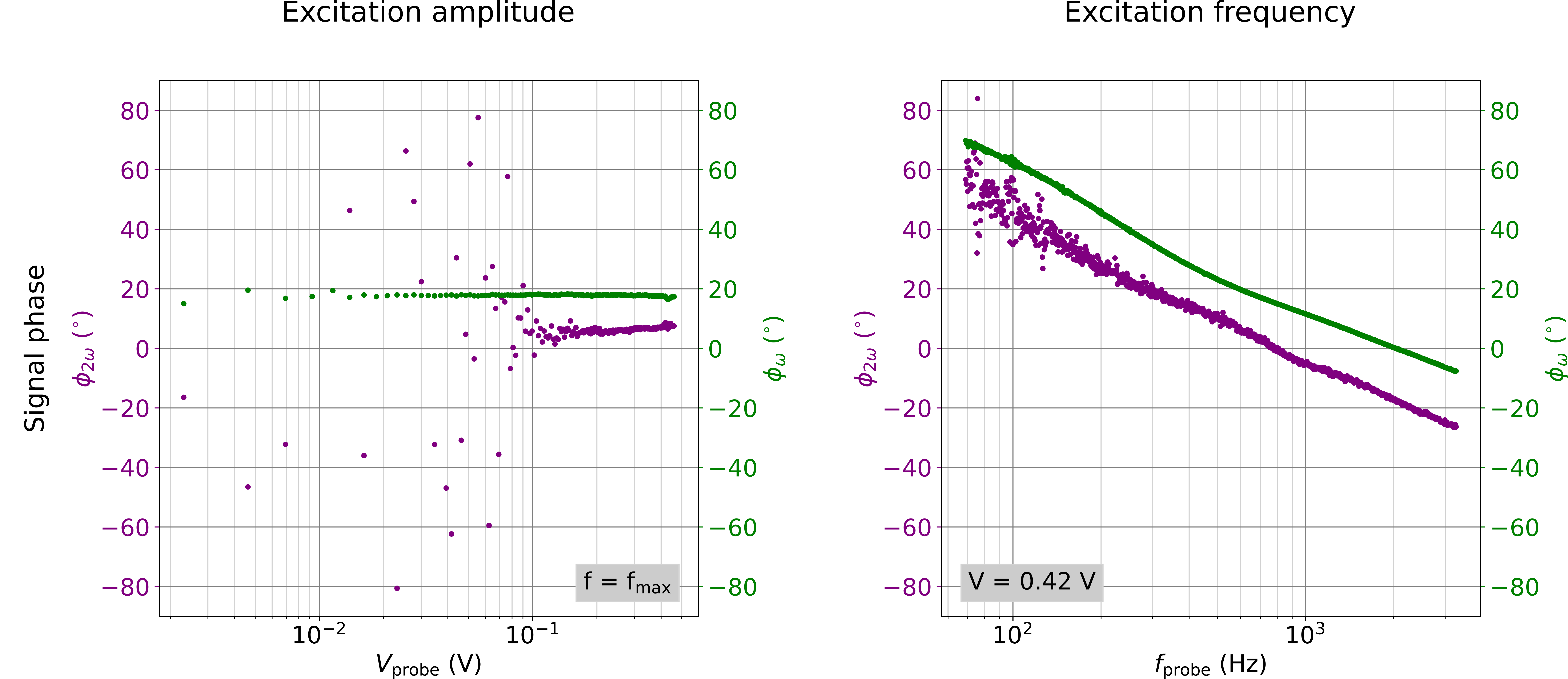}
    \caption{Parameter study on the dependence of the pyroelectric signal, shown in purple, and the electromechanical response, shown alongside in green. \textbf{a} Signal phases with respect to the excitation voltage $V_\text{probe}$. \textbf{b} Signal phases with respect to the excitation frequency $f_\text{probe}$.}
    \label{fig:parametric_phase}
\end{figure*}

The phase counterpart of the parameter study, shown in the main text at Figure~2, are displayed at Figure~\ref{fig:parametric_phase}. The phase is shown to be insensitive to the excitation amplitudes and to depend linearly on the excitation frequency. This shows a constant group velocity and supports single origins for the measured signals.

\clearpage

\subsection{Discussion on contact artefacts}

To exclude the contribution of any contact artefacts arising from the the method to the final signal, we performed three extra analyses each excluding some types of contact artefacts.

First, as shown in the manuscript, a signal at the first harmonic also arises. The parametric studies with amplitude and frequency are presented at Figure~2 in the main text. The mapped signal, shown on Figure~3 in the main text, shows no contrast over the top electrode. The contour of the CIPS flake can be distinguished as substantial bending happens in SThM probe while scanning over sharp edges like those. In addition, the linear relationship of the electrostatic signal $V_{\omega}$ further suggests a mechanical interaction with the surface which would be mediated by the electric field, scaling linearly with the applied voltage on the probe.

\begin{figure}[h!]
    \centering
    \includegraphics[width=0.95\linewidth]{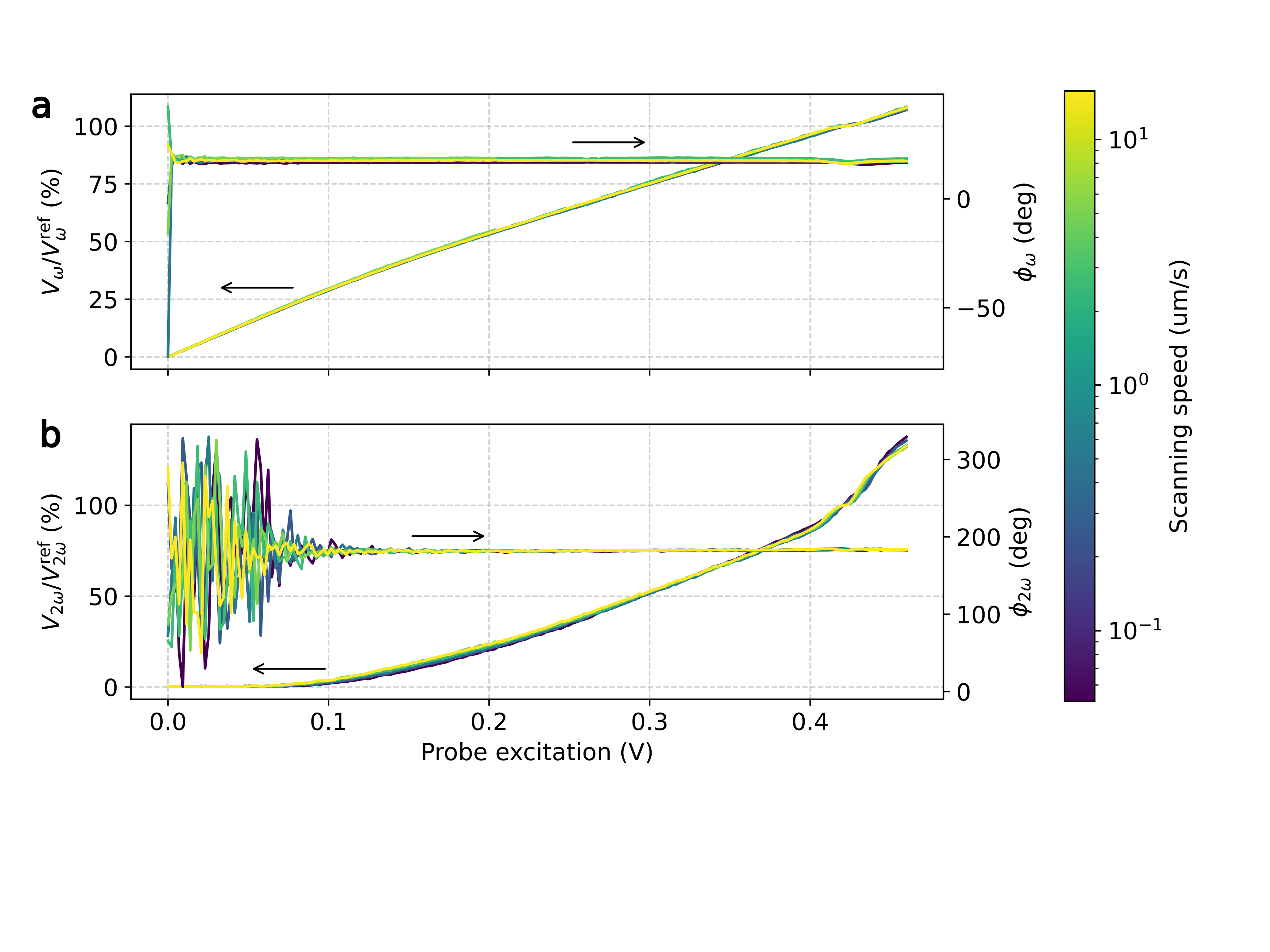}
    \caption{\textbf{a} and \textbf{b} Normalized amplitude and phase measured on the graphene's electrodes at respectively the first and the second harmonic of the excitation. Amplitudes have been normalized by the response at 0.42~V as used in the main text. The color scale denotes the scanning speed, in $\mu$m/s.}
    \label{fig:si_speed}
\end{figure}

We have also investigated DC contact artefacts that would be created by the friction force applied by the tip onto the top electrode. This thermal-mechanical effect of the passage of the probe on a 2D material, sometimes referred to as puckering, has been suggested to strongly affect the tip-sample's thermal contact.\cite{xu_dynamic_2018} This effect should be proportional to the scanning speed of the tip on the surface. To exclude this possibility, we have defined a 1$\mu$m large scanning window which only spanned over the overlapping region. We have performed the same parametric studies as the one presented in the main text for different scanning speed. The results are presented at Figure~\ref{fig:si_speed}. The amplitude values have been normalized to the averaged amplitude measured over this same area during the scan presented at Figure~2 in the main text, acquired at a speed of 6~$\mu$m/s. While varying the scanning speed over two orders of magnitude, no impact could be observed on either the electrostatic or the pyroelectric signal.

\begin{figure}[h!]
    \centering
    \includegraphics[width=0.95\linewidth]{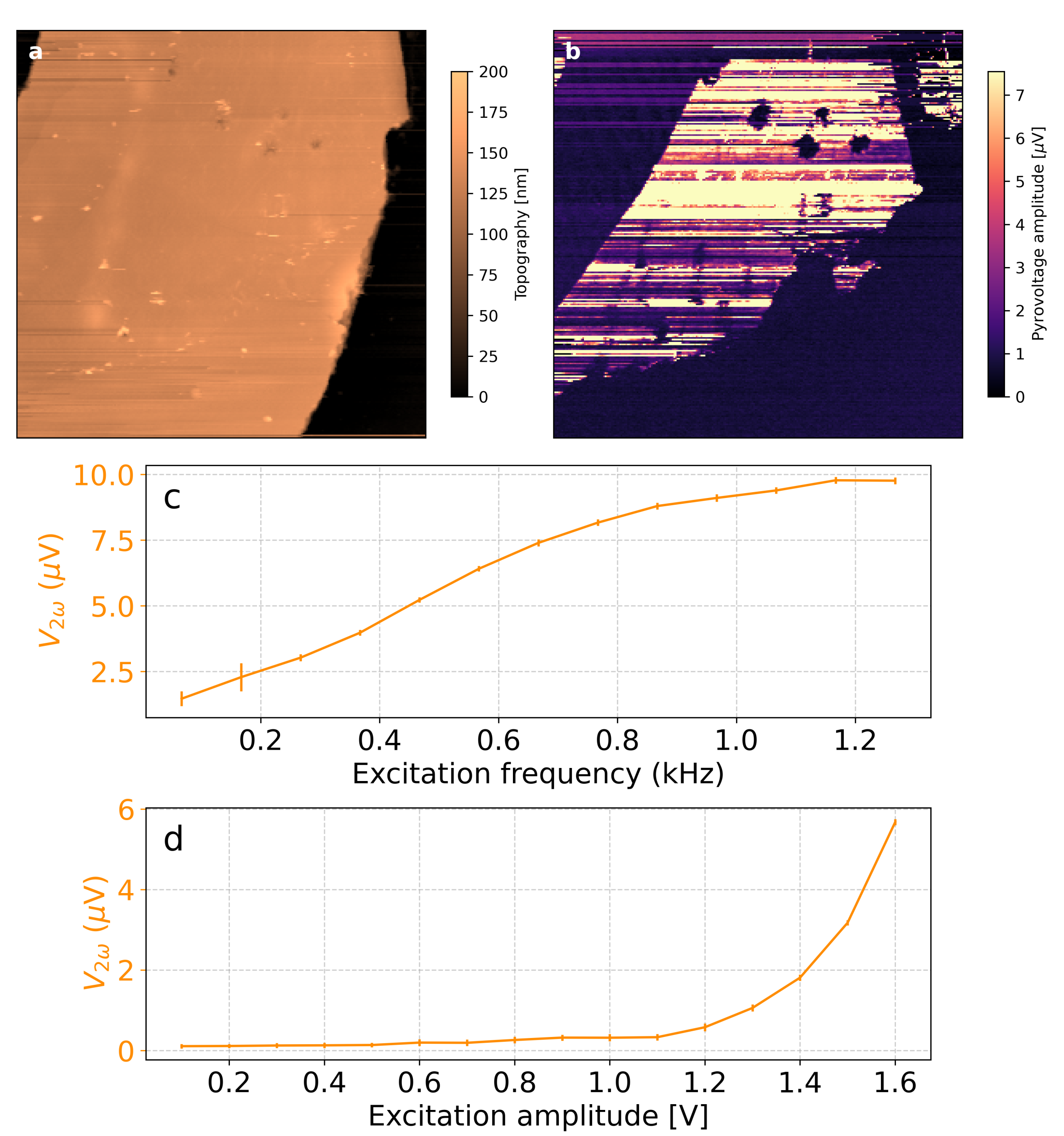}
    \caption{Experiment reproduced inside a high vacuum SThM set-up using a doped silicon SThM probe. \textbf{a} Topography of the graphene/CIPs/graphene stack on the overlapped region. \textbf{b} Amplitude of the pyrovoltage measured between the electrodes. \textbf{c} Measured pyrovoltage amplitude versus the excitation frequency applied on the probe. \textbf{d} Measured pyrovoltage amplitude versus the excitation voltage applied on the probe.}
    \label{fig:si_lancaster}
\end{figure}

The effect of the water meniscus on SThM imaging has been vastly discussed in literature.\cite{kim_ultra-high_2012,mcclelland_signal_2025,spiece_improving_2018} To exclude any water-mediated artefact, we repeated the experiment in a high-vacuum SThM set-up. In addition, a doped silicon SThM probe was used as it allows for a higher heating power (Bruker VITA-HE-NANOTA-200). The obtained results are presented at Figure~\ref{fig:si_lancaster}. While a direct comparison with the results presented in the main text would be difficult as the heat propagation is very different due to the absence of air convection and the modified thermal contact due to the different types of probes, the experiment still yields a strong contrast between the overlapped region and the rest of the scanned region. The quadratic dependency and the high frequency plateau identified in the main text are also found again in the parametric studies presented at Figure~\ref{fig:si_lancaster}c and d. The qualitative coherence of these results with the main study further asserts the reliability of the method and its reproducibility in very different environments.

Finally we have verified the absence of current leakage from the probe to the electrodes by connecting the top electrode to a low-noise current pre-amplifier (Femto DLPCA-200) and applying a DC voltage of \SI{0.5}{\volt} on both ends of the probe. The leakage current was below the detection range of the amplifier (<\SI{10}{\pico\ampere}) and thus cannot realistically play any role in the pyroelectric measurement presented in the main text.

\clearpage

\clearpage
\subsection{Analysis of the RC circuit}
\begin{figure}
    \centering
    \includegraphics[width=0.8\linewidth]{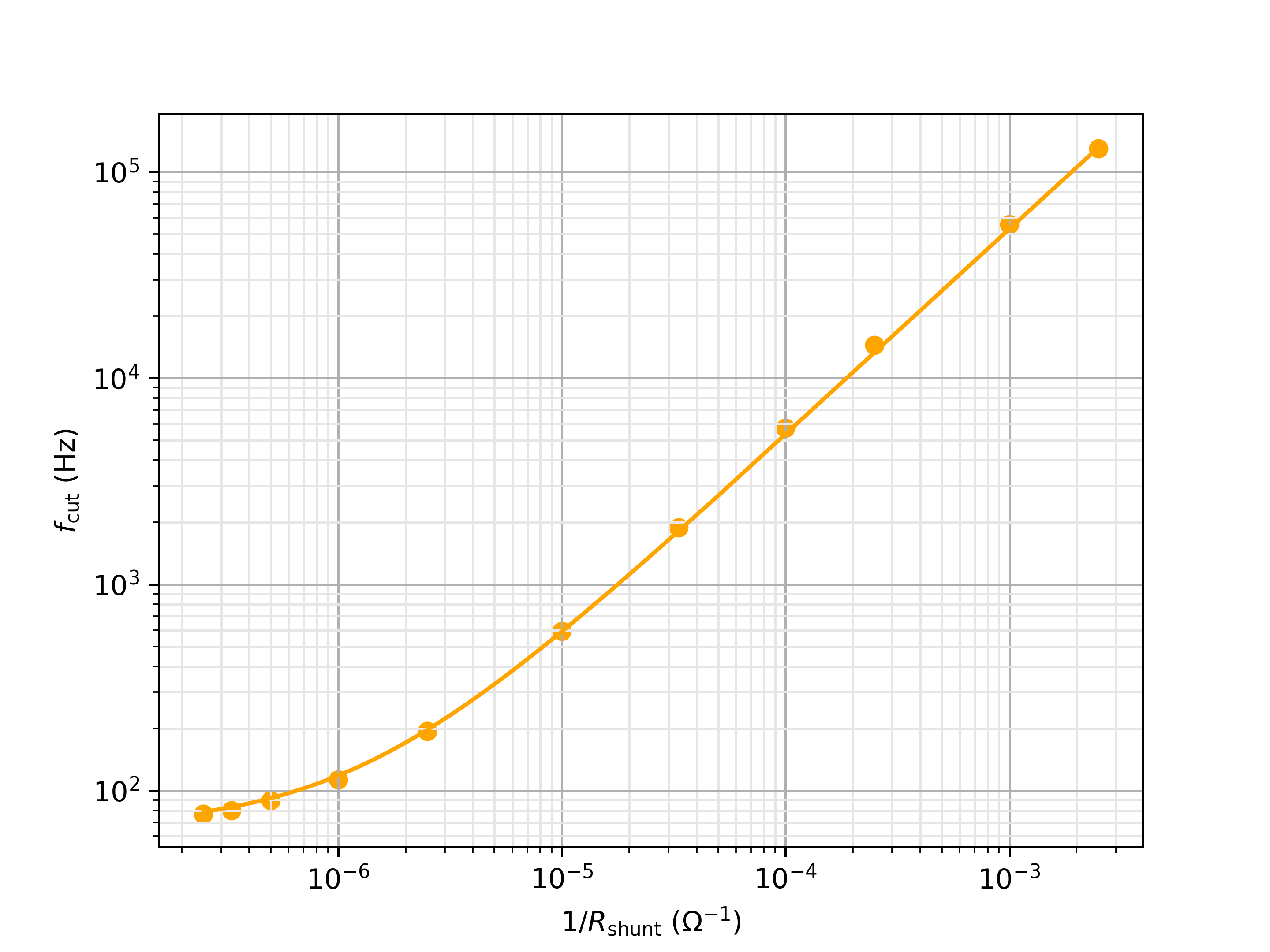}
    \caption{Cut-off frequency of the pyroelectric generator shown with respect to the value of a shunt resistor. In solid line, the R-C filter cut-off frequency's formula with a parallel resistance is shown.}
    \label{fig:si_rc}
\end{figure}

As pointed out in the literature, pyroelectric generator should be modelled as an inner resistance $R_{\mathrm{in}}$ in parallel with a capacitance to account for internal losses and obtain a trustworthy pyroelectric efficiency figure. \cite{zhang_impact_2020} To extract the resistance and capacitance of the pyroelectric device, we include a controlled shunt resistor and measure the cut-off frequency of the device for various shunt resistances. This allows to obtain the RC values independently in a non-invasive manner. The cut-off frequencies are shown on Figure~S\ref{fig:si_rc}. The fitting of Equation~S\ref{eq:rc} is shown as well. This analysis yields an inner resistance of 800~k$\Omega$ and a capacitance of 3.03~nF for the pyroelectric nanogenerator shown in the main text.
\begin{equation}
    f_{cut} \; = \; \frac{1}{2\pi C R_{\mathrm{eq}}} \quad ; \quad R_{\mathrm{eq}} \; = \; \Big(\frac{1}{R_{\mathrm{shunt}}} + \frac{1}{R_{\mathrm{in}}}\Big)^{-1}
    \label{eq:rc}
\end{equation} 

The governing equation for a resistor in parallel with a capacitance is the following :
\begin{equation}
    C\frac{dV}{dt} + \frac{V}{R_{\mathrm{in}}} \;=\; I  \;=\; pA_{\mathrm{tot}}\frac{dT}{dt}
\end{equation}
The frequency domain solution yields the following solution for the amplitude of the generated pyroelectric voltage across the generator $V_0$ :
\begin{equation}
    V_0 \; = \; \frac{pA_{\mathrm{tot}}\Delta T \omega R_{\mathrm{in}}}{\sqrt{1 + (\omega R_{\mathrm{in}} C)^2}}
\end{equation}
This yields a pyroelectric coefficient $p$ of the following form.
\begin{equation}
    p = \frac{|V_{\mathrm{pyro}}|\sqrt{1 + (\omega R_{\mathrm{in}} C)^2}}{A_{\mathrm{tot}}\Delta T \omega R_{\mathrm{in}}}
\end{equation}
This expression directly sets a limit on the low frequency performance of the device, as illustrated in the main text. For an excitation frequency of 769~Hz as reported in the main text, $\omega R_{\mathrm{in}}C$ = 11.71 which justifies using the simplified expression for teh high frequency range, written as follows :

\begin{equation}
    p \approx \frac{|V_{\mathrm{pyro}}|C}{A_{\mathrm{tot}}\Delta T}
\end{equation}

This high-frequency approximation is used hereafter and in the main text, for simplicity.

\clearpage
\subsection{FEA simulations}

We used COMSOL MUltiphysics to perform finite element simulations of our system. First, we used stationary models to obtain the SThM-generated temperature rise in the CIPS layer. Then, we turned to time-dependent models to compute the pyrovoltage and pyrocurrent and compare it to our results. 

Our model geometry is composed of a 300 nm silicon oxide substrate, a 5 nm bottom graphene electrode, a 220 nm CIPS layer and a 8 nm top graphene electrode. On the top electrode, we define a 50 nm radius boundary region acting as a heat source, similar to the SThM tip. 

The main parameters required for the models are summarized in Table \ref{tab:placeholder}.

\begin{table}[h!]
    \centering
    \begin{tabular}{|c|c|c|c|}
    \hline
      & Silicon oxide & Graphite & CIPS \\
      \hline
      Thermal conductivity (W\,m$^{-1}$\,K$^{-1}$) & 1.4 & {500,500,5} & {2,2,0.2} \\
      Heat capacity (J\,kg$^{-1}$\,K$^{-1}$) & 730 & 710 & 500 \\
      Relative permittivity (-) & 4.2 & 1 & 10 \\
      Electrical conductivity (S\,m$^{-1}$) & 0 & $3\times 10^3$ &  \\
      Pyroelectric coefficient (C\,m$^{-2}$\,K$^{-1}$) & 0 & 0 & $5\times 10^{-5}$ \\
      \hline
    \end{tabular}
    \caption{Main parameters used in the COMSOL models}
    \label{tab:placeholder}
\end{table}

\subsubsection{Modeling CIPS temperature rise}

\paragraph{Extraction of the Pyroelectric Coefficient under Localized Heating}

In the present experiment, the SThM probe generates a strongly nonuniform temperature field in the CIPS layer. To extract the pyroelectric coefficient from the measured open-circuit voltage, we begin from the local constitutive relation for the polarization change

\begin{equation}
\Delta P_z(x,y,z) = p\,\Delta T(x,y,z),
\end{equation}

where $p$ is the out-of-plane pyroelectric coefficient and $\Delta T(x,y,z)$ is the local temperature modulation obtained from finite-element thermal simulations.

For a parallel-plate capacitor with metallic electrodes, the electric potential is uniform across each electrode. Neglecting fringing fields, the electric field inside the CIPS layer can therefore be approximated as spatially uniform and directed along the out-of-plane direction,

\begin{equation}
E_z = -\frac{V_{\mathrm{pyro}}}{t},
\end{equation}

where $V_{\mathrm{pyro}}$ is the pyroelectric voltage and $t$ is the thickness of the CIPS layer.

The electric displacement field is then

\begin{equation}
D_z(x,y,z) = \epsilon_0 \epsilon_r E_z + \Delta P_z(x,y,z).
\end{equation}

Under open-circuit conditions, no net free charge can flow between the electrodes. The total change in free charge on the electrodes must therefore remain zero, which imposes the condition

\begin{equation}
\iint_{A_{\mathrm{tot}}} D_z(x,y)\, dA = 0,
\end{equation}

where $A_{\mathrm{tot}}$ is the electrode overlap area.

To account for the nonuniform temperature distribution through the thickness of the CIPS layer, we introduce the thickness-averaged temperature modulation

\begin{equation}
\overline{\Delta T}(x,y)
=
\frac{1}{t}\int_0^t \Delta T(x,y,z)\, dz .
\end{equation}

Using $\Delta P_z = p\,\overline{\Delta T}(x,y)$, the open-circuit condition becomes

\begin{equation}
\iint_{A_{\mathrm{tot}}}
\left(
\epsilon_0 \epsilon_r E_z + p\,\overline{\Delta T}(x,y)
\right)\, dA = 0 .
\end{equation}

Since $E_z$ is uniform across the capacitor area,

\begin{equation}
\epsilon_0 \epsilon_r A_{\mathrm{tot}} E_z
+
p \iint_{A_{\mathrm{tot}}} \overline{\Delta T}(x,y)\, dA
= 0 .
\end{equation}

Substituting $E_z = -V_{\mathrm{pyro}}/t$ yields

\begin{equation}
V_{\mathrm{pyro}}
=
\frac{p\,t}{\epsilon_0 \epsilon_r A_{\mathrm{tot}}}
\iint_{A_{\mathrm{tot}}} \overline{\Delta T}(x,y)\, dA .
\end{equation}

Using the definition of the thickness-averaged temperature modulation, the area integral can be rewritten as a volume integral over the CIPS layer,

\begin{equation}
t \iint_{A_{\mathrm{tot}}} \overline{\Delta T}(x,y)\, dA
=
\iiint_{\mathrm{CIPS}} \Delta T(x,y,z)\, dV .
\end{equation}

Therefore, the pyroelectric voltage may be expressed directly in terms of the volume-integrated temperature field,

\begin{equation}
V_{\mathrm{pyro}}
=
\frac{p}{\epsilon_0 \epsilon_r A_{\mathrm{tot}}}
\iiint_{\mathrm{CIPS}} \Delta T(x,y,z)\, dV .
\end{equation}

Solving for the pyroelectric coefficient gives

\begin{equation}
p
=
\epsilon_0 \epsilon_r A_{\mathrm{tot}}
\frac{|V_{\mathrm{pyro}}|}
{\iiint_{\mathrm{CIPS}} \Delta T(x,y,z)\, dV}.
\end{equation}

This expression is used to extract the pyroelectric coefficient from the measured open-circuit voltage and the simulated temperature field obtained from the finite-element thermal model.

\paragraph{Relation to an Effective Heated-Area Approximation}

For intuitive comparison with a locally heated capacitor picture, one may define an effective heated area $A_{\mathrm{heat}}$ and an effective temperature modulation $\Delta T_{\mathrm{eff}}$ such that

\begin{equation}
\iiint_{\mathrm{CIPS}} \Delta T(x,y,z)\, dV
=
t\,A_{\mathrm{heat}}\,\Delta T_{\mathrm{eff}} .
\end{equation}

The pyroelectric voltage then becomes

\begin{equation}
V_{\mathrm{pyro}}
=
\frac{p\,t}{\epsilon_0 \epsilon_r A_{\mathrm{tot}}}
A_{\mathrm{heat}} \Delta T_{\mathrm{eff}},
\end{equation}

and therefore

\begin{equation}
p
=
\frac{\epsilon_0 \epsilon_r}{t}
\frac{A_{\mathrm{tot}}}{A_{\mathrm{heat}}}
\frac{|V_{\mathrm{pyro}}|}{|\Delta T_{\mathrm{eff}}|}.
\end{equation}

This effective heated-area form is mathematically equivalent to the volume-integral formulation above but depends on the specific definition of $A_{\mathrm{heat}}$ and $\Delta T_{\mathrm{eff}}$. In the present work, the volume-integral approach is preferred because it directly uses the simulated three-dimensional temperature field and avoids introducing additional geometrical approximations.

\paragraph{Finite-Element Thermal Modeling}

To better understand the heat spreading in the device, we compute the stationary temperature field generated by a localized heat source representing the SThM tip. The simulations assume a circular contact with a radius of \SI{50}{nm} and a temperature offset of \SI{1}{K} applied at the tip–sample interface.

Two configurations were considered: (i) the tip in contact with the top graphene electrode and (ii) the tip in direct contact with the CIPS layer.

\begin{figure}
    \centering
    \includegraphics[width=0.75\linewidth]{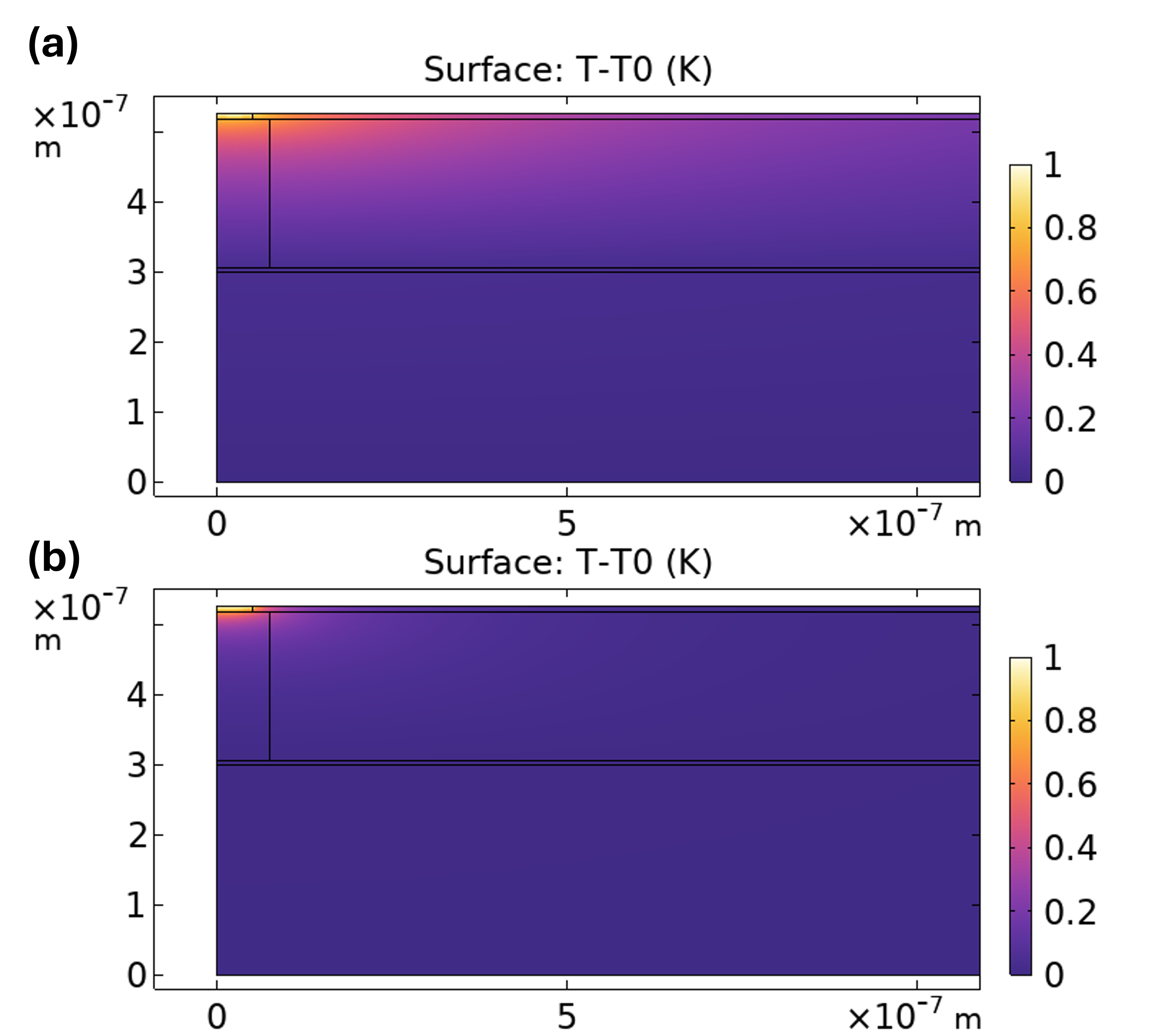}
    \caption{Finite-element simulations of the temperature field generated by a localized heat source (50 nm radius, 1 K temperature offset). (a) Tip in contact with the top graphene electrode. (b) Tip in direct contact with the CIPS layer.}
    \label{fig:COMSOLheat}
\end{figure}

Figure~\ref{fig:COMSOLheat} shows the simulated temperature distributions for both cases. When the tip contacts the graphene electrode, the high in-plane thermal conductivity of graphene leads to significant lateral heat spreading. In this configuration the temperature rise propagates laterally over several hundred nanometers, while the bottom electrode experiences only a negligible temperature increase.

In contrast, when the tip is in direct contact with the CIPS layer, the much lower thermal conductivity of CIPS confines the temperature rise to the immediate vicinity of the tip. The lateral temperature spreading is therefore strongly suppressed.

These two configurations lead to significantly different values of the volume-integrated temperature in the CIPS layer. For the case where the tip contacts the graphene electrode, the integral $\iiint \Delta T\, dV$ equals $8.28 \times 10^{-19}$~m$^{3}$K, whereas for direct contact with CIPS the value is $2.26 \times 10^{-20}$~m$^{3}$K. This large difference highlights the important role played by the graphene electrode in redistributing heat laterally across the device.

\clearpage

\clearpage
\section*{Bibliography}
\bibliography{mybib}{}
\bibliographystyle{MSP}